\documentclass[pra,aps,showpacs]{revtex4}
\usepackage{amsthm}
\usepackage{amsmath}
\usepackage{epsfig}
\def\trace#1{{\mathrm{Tr}\left[#1\right]}}
\def\abs#1{{\vert #1 \vert}}

\def\id{\mathcal{I}}
\def\ket#1{\left\vert #1 \right\rangle}
\def\tensorm{\otimes}
\def\comm#1#2{\left[#1,#2\right]}

\def\bra#1{\left\langle #1 \right\vert}
\def\ensavg#1{\left\langle #1 \right\rangle}
\def\v#1{{\bf #1}}
\def\refsec#1{Sec.\ \ref{Section::#1}}

\def\refeqn#1{Eq.\ (\ref{Equation::#1})}

\def\refeqs#1#2{Eqs.\ (\ref{Equation::#1}) and (\ref{Equation::#2})}
\def\refeqto#1#2{Eqs.\ (\ref{Equation::#1}--\ref{Equation::#2})}

\begin{document}
\title{Universal Scaling of Hyperfine-Induced Electron Spin Echo Decay}

\author{Neil Shenvi}
\author{Rogerio de Sousa}
\author{K. Birgitta Whaley}
\affiliation{Department of Chemistry and the Kenneth
S. Pitzer Center for Theoretical Chemistry, University of California,
Berkeley,
        Berkeley, CA 94720}

\date{\today}

\begin{abstract}
The decoherence of a localized electron spin in a lattice of nuclear 
spins is an important problem for potential solid-state implementations of 
a quantum computer.  We demonstrate that even at high fields, virtual 
electron spin-flip processes due solely to the hyperfine interaction
can lead to complex nuclear spin dynamics.  These dynamics, in turn, can 
lead to single electron spin phase fluctuation and decoherence.
We show here that remarkably, a spin echo pulse sequence can almost 
completely reverse these nuclear dynamics except for a small visibility loss,
thereby suppressing contribution of the hyperfine interaction to $T_2$ 
processes.  For small systems, we present numerical evidence which 
demonstrates a universal scaling of the magnitude of visibility loss
that depends only on the inhomogeneous line width of the system and the
magnetic field.
\end{abstract}


\maketitle
\section{Introduction} 
\label{Section::Introduction}
The hyperfine coupling between an excess electron spin and its surrounding nuclear 
spin lattice is an important source of decoherence in spin-based 
quantum dot schemes for solid state quantum computation 
\cite{Loss:98A,Khaetskii:03L}.  Due to the hyperfine coupling, the Zeeman 
energy of the electron spin can be exchanged with the Overhauser energy of the 
nuclear lattice, resulting in loss of longitudinal electron spin 
polarization.  In addition, time fluctuations of the Overhauser field lead to 
loss of coherence of the electronic spin state\cite{deSousa:03B1}.  In 
a Si based quantum computer, it may be possible to avoid this interaction 
by using isotopically purified Si ($^{28}$Si has no nuclear spin) 
\cite{deSousa:03B1,Yablonovitch:03}.  However, 
for many III-V semiconductors such as GaAs, there are no spin-zero nuclear 
isotopes; therefore the influence of the hyperfine coupling will have to 
be taken into account.  A comprehensive understanding of the hyperfine 
interaction may allow the loss of electron spin coherence to be 
minimized through intelligent hardware design or pulse sequence 
engineering.

Numerous studies have investigated the electron spin decay induced by 
the hyperfine interaction under free evolution.  
At zero external field, calculations of the exact quantum dynamics for small 
systems established that the hyperfine interaction generates substantial 
entanglement between the electron and the nuclear spin lattice 
\cite{Schliemann:02B}.  
At low external fields, analytic solutions were obtained
to describe the decay of electron spin correlation functions, finding 
that longitudinal spin relaxation occurs via 
power law decay on a timescale governed by $A\sqrt{N}$ where $N$ is the number of nuclei 
and $A$ is the average hyperfine coupling strength\cite{Khaetskii:03B,Khaetskii:03L}.  
In \cite{Privman:02}, decoherence due to the hyperfine Hamiltonian was studied using
a Markovian approximation to nuclear dynamics.  A generalized master equation 
has also been used to describe non-Markovian dynamics in both 
the high- and low-field regimes at short timescales, again finding a power 
law decay of coherence\cite{Coish:04B}.  However, semi-classical calculations
have shown that correlation functions exhibit long timescale, oscillatory behavior 
rather than chaotic dynamics, suggesting that correlation function decay is due to 
averaging over initial conditions rather than from inherent system ergodicity 
\cite{Erlingsson:04}.  In all of these previous 
publications, electron spin decay was studied under free evolution.  However, 
in most experiments \cite{Tyryshkin:03B, Abe:04B}
and in the context of controlled spin dynamics for quantum computation, it 
is desirable to measure coherence in the context of a spin echo pulse 
sequence.  Therefore, in this 
paper, we will investigate electron spin coherence and nuclear spin dynamics 
due to the hyperfine coupling under an idealized spin echo experiment.  

In physically relevant systems, other interactions will also contribute to electron
spin decoherence.  For instance, decoherence of the electron spin due to
spin-orbit induced coupling to phonons was studied in \cite{Mozyrsky:02B,Semenov:03L}.
Spectral diffusion arising from dipolar nuclear-nuclear interaction has been
addressed in \cite{deSousa:04X}, where it was shown to be effectively controllable
with suitably designed CPMG pulse sequences.
Low energy effective Hamiltonians describing a central spin coupled to a spin bath 
have been analyzed with operator instanton methods \cite{Stamp:00}. 
The effect of precession of bath spins on coherence of a central spin has also been 
addressed in the spin bath models,
and distinguished from the
decoherence effects caused by `co-flips' of bath spins with the central
spin \cite{Dube:01}.
Because we are interested specifically in the effect of the hyperfine
interaction here, we will neglect other effects such as phonon coupling or dipolar 
nuclear-nuclear coupling which can also contribute to decoherence, in order to 
isolate the effect of the hyperfine interaction \cite{Shenvi:04X1}.

Rather than resort to approximation methods, we simulate the full 
electron-nuclei system through exact diagonalization.  
Due to the exponential growth of the 
Hilbert space, only small numbers of nuclear spins can be treated using 
this method.  However, even for small systems we observe complex dynamics 
which will serve to illustrate the fundamental physics for larger systems.
We focus mainly on the regime 
of high magnetic fields in which the external magnetic field is 
greater than the total Overhauser field of the nuclei.  
This regime is both experimentally realizable and practically desirable for 
applications in quantum computation, since single spin measurement requires a 
high effective magnetic field \cite{Hanson:03L}.  In this high field 
regime, we 
obtain several non-intuitive results that have implications for 
electron spin coherence.  
First, we find that even at high magnetic fields, rich nuclear 
dynamics can still occur.  Because high magnetic fields suppress 
single flip-flops between the electron and a nucleus, it might be 
naively assumed that all dynamics are suppressed.  However, virtual flip-flops 
enable nuclear dynamics to persist even at high fields.  Second, 
as expected, we observe rapid decay of the in-plane magnetization of the 
electron under free evolution of the system (no spin echo), both for a pure 
initial state 
and for a completely mixed initial state.
However, if we simulate a spin echo sequence, we find that the 
complex nuclear dynamics are almost completely reversed and the in-plane 
magnetization of the electron is almost completely recovered, except for a 
small visibility decay.  This 
phenomenon, if it can be 
generalized to larger numbers of spins, would imply that the spin echo 
pulse sequence can remove almost all electron decoherence caused by the 
hyperfine interaction.  This behavior appears to reflect some partial hidden
symmetry of the hyperfine Hamiltonian in the presence of an external 
field.

This paper is organized as follows: Section II defines our system 
Hamiltonian and gives background information regarding the spin echo 
experiment and the origin of nuclear dynamics.  Section III 
presents our numerical results.  Conclusions are presented in Section 
IV together with discussion.

\section{Background}
\label{Section::Background}
In this section, we present general background information regarding 
our spin system.  First, we will define our Hamiltonian and identify some 
of the symmetries we can use to simplify it.  We also give a brief 
synopsis of decoherence processes and cite how they relate to our system.  
Second, we outline the spin echo experiment and discuss its utility for
removing inhomogeneous contributions to the electron spin decay.  Finally, 
we derive an effective Hamiltonian 
which explains the origin of the nuclear dynamics that we observe in our 
numerical simulations.

The system we study is that of a localized electron
with spin operator $\v{S}$ in a lattice of $N$ nuclear spins, with spin 
operators $\v{I}_1, \v{I}_2, \ldots, \v{I}_N$.  Here we consider spin 
$1/2$ nuclei such that $S = I = 1/2$.  The system in question 
could be an impurity electron bound to a doping atom in a semiconductor 
lattice, or an excess electron in a quantum dot.  
The electron spin is coupled to the $j^{th}$ nuclear 
spin via the hyperfine interaction $A_j$.
The Hamiltonian for this system is 
\cite{Khaetskii:03L,Coish:04B,Schliemann:02B,Shenvi:04X1}
\begin{equation} \label{Equation::HFull}
H = \gamma_S B S_z + \gamma_I B \sum_j{I_{jz}} +
\sum_j{A_j \v{S}\cdot\v{I}_j},
\end{equation}
where $\gamma_S$ and $\gamma_I$ are the gyromagnetic ratios of the 
electron and nuclei, respectively, and $B$ is the external magnetic 
field.  Here, we have neglected nuclear-nuclear dipolar coupling, which is known to 
contribute to electron spin decoherence through nuclear 
spectral diffusion, because we would like to isolate the contribution of the 
hyperfine coupling \cite{deSousa:03B2,Shenvi:04X1}.  
The Hamiltonian in \refeqn{HFull} conserves the total spin angular momentum.
Thus, we can immediately block diagonalize the Hamiltonian based on the 
$J_z = S_z + \sum_j{I_{jz}}$ operator.
For convenience, we will label each of these blocks by the quantum number 
$L$ where $L$ is the total number of ``down'' spins (i.e. $J_z = \hbar(N+1-2L)/2$).
  For a given block with quantum number $L$, we can then remove 
the nuclear Zeeman energy from the Hamiltonian by subtracting an overall 
constant,
\begin{equation} \label{Equation::HFullJ}
H = (\gamma_S-\gamma_I) B S_z + E_L + \sum_j{A_j \v{S}\cdot\v{I}_j},
\end{equation}
where
\begin{equation}
E_L = \hbar \gamma_I B (N-2L+1)/2.
\end{equation}
Since each block can be treated independently, all further analysis 
pertains to the subspace specified by some given value of the quantum 
number $L$.  In general, we will also drop $E_L$ since it will add only an 
overall phase to the evolution within any given subspace.  

The full Hamiltonian can be separated into a 
zeroth order Hamiltonian $H_0$ and a perturbation $V$,
\begin{eqnarray}
H &=& H_0 + V \\
H_0 &=& (\gamma_S-\gamma_I) B S_z + \sum_j{A_j 
S_z   I_{jz}} \\
V &=& \frac{1}{2}\sum_j{A_j \left(S_+ I_{j-}+S_- I_{j+}\right)}.
\end{eqnarray}
At high $B$ fields, this separation can be used to gain insight 
into the origin of system dynamics \cite{Shenvi:04X1}; however, it should be 
emphasized that 
we simulate here the full quantum dynamics using exact diagonalization, 
not by using any type of perturbative treatment.
We define $\ket{\Uparrow}$ and $\ket{\Downarrow}$ to represent the $+z$ 
and $-z$ polarized electron states, respectively.  We also define
$\v{z}$ to be an $N$-bit string of $+1$'s and $-1$'s representing a given 
state of the nuclei in the $z$ basis.  
The eigenstates of the unperturbed Hamiltonian, $H_0$, are then given by
$\ket{\Uparrow,\v{z}}$ or $\ket{\Downarrow,\v{z}}$.

In such a system, the electron spin undergoes decoherence processes, which 
are usually described by the Bloch equations and which are usually 
characterized by coherence timescales $T_1$ and $T_2$.  However, the Bloch 
equations assume an exponential form for coherence decay which is not a valid 
assumption for hyperfine-induced decay \cite{Khaetskii:03L,Coish:04B}.  If we 
wish to characterize the timescale of decoherence, we must select a 
suitable definition for $T_1$ and $T_2$.  In this paper, we define $T_1$ to be 
the time it takes for the longitudinal electron spin magnetization $S_z(t)$ to 
decay to $1/e$ of its initial value.  Similarly, we define $T_2^*$ to be the 
time it takes for the magnitude of the in-plane magnetization $S_+(t)$ to 
decay to $1/e$ of its initial value.  $T_2^*$ contains contributions from 
inhomogeneous broadening, which results either from the measurement of an 
ensemble of spins in different local environments, 
or from the measurement of a single spin in a mixed initial 
state, $\rho(0)$.  Therefore we also define the spin echo coherence time 
$T_2$, which is defined as the time it takes for the 
spin echo envelope to decay to $1/e$ of its initial value \cite{Hahn:50R}.  
$T_2$ contains no contributions from inhomogeneous broadening, which is 
removed by the spin echo pulse sequence.  These definitions are equivalent to 
the standard definitions in the case of exponential decay.  Throughout this paper, 
we will take ``decoherence" to refer to all processes which lead to a loss of 
electron spin polarization, both longitudinal and transverse.  ``Dephasing" will 
refer to processes which lead to a loss of transverse polarization only.

We now note that due to the large Zeeman energy of the electron, 
there will be an energy gap between the manifold of electron spin-up 
and electron spin-down eigenstates of $H_0$.  
It is clear that for large enough fields, $T_1$ will be infinite because 
there is then no efficient mechanism for the longitudinal relaxation of 
the electron spin.  If we define the total Overhauser field,
\begin{equation}
B_c = \frac{\hbar}{\gamma_S-\gamma_I}\sum_j{A_j}
\end{equation}
then when $B > B_c$ we can derive several 
results\cite{Shenvi:04X1}:
\begin{enumerate}
\item The $S_z$ label is approximately a good 
label for the eigenstates of $H$.  In other words, eigenstates of $H$ 
are either nearly electron spin-up ($\ket{\Uparrow}$) or nearly electron 
spin-down ($\ket{\Downarrow}$).  We will 
refer to these states as ``+" and "-" eigenstates, respectively.
\item There is an energy gap of approximately $\hbar 
\gamma_S B$ between nearly spin-up (``+") eigenstates of $H$ 
and 
nearly spin-down (``-") eigenstates of $H$.
\item Because $S_z$ is nearly a good label for the eigenstates of $H$, 
$S_z(t)$ can only deviate a small amount from its initial value,  
\begin{equation}
\abs{S_z(t) - S_z(0)} \leq {\mathcal O}\left(\frac{B_c^2}{B^2}\right).
\end{equation}
\end{enumerate}
Hence, we can identify $B_c$ as a critical field for the longitudinal 
relaxation of the system.  (Results 1-3 are proved rigorously in 
\cite{Shenvi:04X1}.)

Although longitudinal relaxation is suppressed at high fields, 
resulting in an infinite value for the $T_1$ coherence time, the 
same is not necessarily true for the intrinsic in-plane coherence time, 
$T_2$.  In fact, numerical evidence (see \refsec{Numerics}) demonstrates 
that $T_2$ is governed by a second critical field, 
$\Delta_c$, defined as,
\begin{equation} \label{Equation::BCPrimeDef}
\Delta_c = \frac{\hbar}{\gamma_S-\gamma_I} \sqrt{\sum_j{A_j^2}}.
\end{equation}
Since we are concerned here with the in-plane relaxation of the 
electron spin, from now on when we speak of the ``critical 
field" we will be referring to $\Delta_c$.  This critical field is equivalent 
to the inhomogeneous broadening linewidth due to the hyperfine interaction
(i.e., $T_2^* \sim 1/(\gamma_S \Delta_c)$).  What is remarkable, however, is 
that this quantity also acts as a critical field for \emph{intrinsic} 
broadening.

This intrinsic broadening ($T_2$) can be caused by nuclear dynamics which 
stem from second-order spin flip processes.  Because these second-order 
processes conserve $S_z$, they are not suppressed as strongly as $T_1$ 
processes.  It is these second-order processes which cause the 
divergences in second-order perturbation theory noted in 
\cite{Khaetskii:03L}.  In addition to intrinsic broadening, 
inhomogeneous broadening which contributes to $T_2^*$ can occur in an 
ensemble measurement when an ensemble of electrons is 
subjected to a slightly inhomogeneous magnetic field.  Inhomogeneous 
broadening can also occur when a single electron experiences an ensemble 
of Overhauser fields due to the nuclei being in a mixed state.  In our 
calculations, inhomogeneous broadening will result from the use of a fully 
mixed initial state for the nuclear spins.  This 
static, inhomogeneous contribution to $T_2^*$ can be contrasted to the 
dephasing induced by dynamic sources, such as a time-varying external 
field or a fluctuating nuclear Overhauser field caused by nuclear 
spin dynamics.  To remove the contribution due to static, 
inhomogeneous broadening and recover the intrinsic coherence time, $T_2$, 
we must perform a spin echo sequence ($\pi/2-\tau-\pi-\tau-$echo).

We now turn our attention to a detailed analysis of the spin echo 
experiment for this system, which will be 
crucial in interpreting our numerical results.  The 
spin echo experiment was developed by Hahn \cite{Hahn:50R} to remove 
inhomogeneous broadening in an ensemble spin measurement.  We will assume 
that at the beginning of the spin echo experiment, the electron spin has 
been rotated to point in the $+x$ direction.  The system is 
then allowed to evolve freely for some time $\tau$.  Next, an idealized 
$\pi$-pulse which flips the spin of the electron is applied.  The $\pi$ 
pulse is described by the operator,
\begin{equation} \label{Equation::PiPulseDef}
R_\pi = \ket{\Uparrow}\bra{\Downarrow} + \ket{\Downarrow}\bra{\Uparrow}.
\end{equation}  
After another period 
of free evolution for time $\tau$, the magnitude of the in-plane 
magnetization of the electron, 
$S_+ = S_x + i S_y$, is measured, yielding a measure of single spin coherence.  

In addition to removing the effects of inhomogeneous broadening, the 
spin echo experiment can remove dephasing 
due to a broad class of Hamiltonians having the form,
\begin{equation} \label{Equation::HE}
H_{se} = \ket{\Uparrow}\bra{\Uparrow} \otimes V_\Uparrow + 
\ket{\Downarrow}\bra{\Downarrow}\otimes V_\Downarrow,
\end{equation}
where $V_\Uparrow$ and $V_\Downarrow$ are arbitrary operators on 
the nuclear spins that have the 
commutation property $\comm{V_\Uparrow}{V_\Downarrow} = 0$.  The in-plane 
magnetization after the spin echo sequence is given by
\begin{equation}
\ensavg{\tilde{S}_+(\tau;\tau)} = \trace{\rho(0)U^\dagger(\tau)R_\pi 
U^\dagger(\tau)S_+ U(\tau)R_\pi U(\tau)},
\end{equation}
where
\begin{equation}
U(\tau) = e^{i H_{se} \tau / \hbar}.
\end{equation}
First we note that we can write $H_{se}$ as
\begin{equation}
H_{se} = \left(V_\Uparrow+V_\Downarrow\right)\id + 
\left(V_\Uparrow-V_\Downarrow\right)S_z.
\end{equation}
Then we note the relations,
\begin{eqnarray}
R_\pi S_z R_\pi &=& -S_z \\
R_\pi S_+ R_\pi &=& S_-.
\end{eqnarray}
Using these facts, we find that
\begin{eqnarray}
\ensavg{\tilde{S}_+(\tau;\tau)} 
&=& \trace{\rho(0)U^\dagger(\tau)R_\pi U^\dagger(\tau)S_+ U(\tau)R_\pi U(\tau)} \\
&=& \trace{\rho(0)
e^{-i (V_\Uparrow-V_\Downarrow)S_z \tau / \hbar}
R_\pi e^{-i (V_\Uparrow-V_\Downarrow)S_z \tau / \hbar}
R_\pi
S_- R_\pi e^{i (V_\Uparrow-V_\Downarrow)S_z \tau / \hbar}
R_\pi e^{i (V_\Uparrow-V_\Downarrow)S_z \tau / \hbar}} 
\\
&=& \trace{\rho(0)
e^{-i (V_\Uparrow-V_\Downarrow)S_z \tau / \hbar}
e^{i (V_\Uparrow-V_\Downarrow)S_z \tau / \hbar}
S_- e^{-i (V_\Uparrow-V_\Downarrow)S_z \tau / \hbar}
e^{i (V_\Uparrow-V_\Downarrow)S_z \tau / \hbar}} \\
&=& \trace{\rho(0)S_-} \\
&=& \ensavg{S_-(0)}. 
\end{eqnarray}
Hence, we arrive at the conclusion that the spin echo experiment 
removes decay of in-plane magnetization for all Hamiltonians of the form 
given in \refeqn{HE}.

To understand the origin of nuclear dynamics in this system, we can 
derive
an effective Hamiltonian which contains nuclear-nuclear 
interactions.  Let 
$\ket{\psi^+}$ be a 
``+" eigenstate of the Hamiltonian, $H$, i.e. $\ket{\psi^+}$ has 
primarily 
electron spin-up character.  
Without loss of generality, $\ket{\psi^+}$ can be written as
\begin{equation}
\ket{\psi^+} = \ket{\Uparrow,\psi_\Uparrow^+} + 
\ket{\Downarrow,\psi_\Downarrow^+}.
\end{equation}
Because the perturbation $V$ flips the polarization of the electron, 
the 
action of $H$ on the electron spin-up and electron spin-down 
subspaces yields the two simultaneous equations,
\begin{eqnarray}\label{Equation::HPsiPlus}
H_0 \ket{\Uparrow,\psi_\Uparrow^+} + V 
\ket{\Downarrow,\psi_\Downarrow^+} 
&=& E_+ \ket{\Uparrow,\psi_\Uparrow^+}\\
\label{Equation::HPsiMinus}
H_0 \ket{\Downarrow,\psi_\Downarrow^+} + V 
\ket{\Uparrow,\psi_\Uparrow^+} 
&=& E_+ \ket{\Downarrow,\psi_\Downarrow^+}.
\end{eqnarray}
\refeqn{HPsiMinus} can be solved for 
$\ket{\Downarrow,\psi_\Downarrow^+}$ 
and
the resulting expression inserted into \refeqn{HPsiPlus} yields
\begin{equation} \label{Equation::WignerPsiPlus}
H_0 \ket{\Uparrow,\psi_\Uparrow^+} + V \frac{1}{E_+ - H_0} V 
\ket{\Uparrow,\psi_\Uparrow^+} = 
E_+ \ket{\Uparrow, \psi_\Uparrow^+}.
\end{equation}
Because of the energy gap between the spin-up and spin-down states, 
the operator $1/(E_+-H_0)$ is always well-defined \cite{Shenvi:04X1}.  
Because the left-hand side of \refeqn{WignerPsiPlus} depends on $E_+$, it is 
not a true Schr\"{o}dinger equation; to obtain $E_+$ exactly, 
\refeqn{WignerPsiPlus} must be solved self-consistently.  
However, if we use $E_+ \approx \hbar(\gamma_S - \gamma_I) B / 2$, then we can 
obtain an effective Hamiltonian from \refeqn{WignerPsiPlus}.
The effective Hamiltonian for the electron spin-up subspace is
\begin{eqnarray} \label{Equation::Heffp}
H_{\mathrm{eff}}^+ &=& H_0 + V_{\mathrm{eff}}^+ \\
\label{Equation::Veffp}
V_{\mathrm{eff}}^+ &=& \frac{\hbar^2}{4}\sum_{j,k}{A_j A_k I_{j-} 
\frac{1}{\hbar(\gamma_S-\gamma_I)B+\frac{1}{2}\hbar\sum_j{A_j I_{jz}}}I_{k+}}.
\end{eqnarray}
We obtain a similar, but not identical, 
effective Hamiltonian for the spin-down 
subspace (note the transposition of the $I_-$ and $I_+$ operators),
\begin{eqnarray} \label{Equation::Heffm}
H_{\mathrm{eff}}^- &=& H_0 + V_{\mathrm{eff}}^- \\
\label{Equation::Veffm}
V_{\mathrm{eff}}^- &=& -\frac{\hbar^2}{4}\sum_{j,k}{A_j A_k I_{j+} 
\frac{1}{\hbar(\gamma_S-\gamma_I)B+\frac{1}{2}\hbar\sum_j{A_j I_{jz}}}I_{k-}}.
\end{eqnarray}
\refeqs{Veffp}{Veffm} show that the overall coupling between nuclei 
does indeed decrease at high fields, because the operator $1/(E-H_0)$ 
scales approximately as $1/B$.  However, the energy cost of flip-flopping two 
nuclei $j$ and $k$ is proportional to $A_j - A_k$.  Thus, if $A_j$ and 
$A_k$ are close in value, the nuclei can flip-flop even at high fields.  

If we now examine the hyperfine Hamiltonian in \refeqn{HFull} in light of 
this discussion, we note that 
it is not immediately clear whether the spin echo experiment 
will remove all dephasing.  $H$ does not have the form of 
$H_{se}$ given in 
\refeqn{HE}.  Furthermore, as noted above, although 
real electron spin-flip processes are suppressed at 
high magnetic field, the electron can instead undergo a virtual 
flip-flop which corresponds to the perturbation $V$ acting twice on the 
initial state.  This action leaves the spin state 
of the electron unchanged, but flip-flops the spins of two nuclei.  
Because this process does not produce a net 
change in the longitudinal polarization of the electron, it can occur even 
at high fields (when $T_1$ processes are substantially suppressed) and hence 
may contribute to electron spin decoherence.  We will provide explicit 
examples of the nuclear spin dynamics resulting from these virtual electron spin flip processes in 
Section III.

\section{Numerics} \label{Section::Numerics}
We now show numerical calculations for one electron and $N = 5-13$ 
nuclei.  Due to the exponential size of the Hilbert space, simulating 
larger systems rapidly becomes unfeasible.  However, even in small 
systems, we will see that the basic physics of the hyperfine interaction 
and its role in the features of the resulting nuclear dynamics become
apparent.  We will show that the spin echo 
envelope decay depends only on the ratio of the external field $B$ to 
the critical field $\Delta_c$. 

We will make use of two different initial states in our simulations.  
The first initial state we consider is a pure state in which the nuclei 
are all pointed in either 
the $+z$ or $-z$ direction:
\begin{equation} \label{Equation::IC1}
\ket{\psi_0} = 
\ket{\Leftarrow}
\tensorm\ket{\v{z}_0},
\end{equation}
where $\ket{\v{z}_0}$ is a randomly chosen simultaneous eigenstate of the 
$I_{jz}$ operators.  Our second initial state is the completely mixed 
nuclear density matrix $\rho(0)$, given by
\begin{equation} \label{Equation::Rho0Def}
\rho(0) = c\ket{\Leftarrow}\bra{\Leftarrow}\tensorm\id_I,
\end{equation}
where $c$ is a normalization constant and $\id_I$ is the identity operator on 
the nuclear spins.
In both cases, the electron is initially polarized in the $+x$ direction.
We simulate the nuclear dynamics for a 
specific set of hyperfine constants, $\{A_j\}$, where $j = 1 \ldots N$ and $N=9$.  The 
hyperfine constants are selected randomly from a uniform distribution on the range 
$\left[.1,.2\right]\, \mathrm{Tesla}$ yielding a total Overhauser field 
of $B_c = 1.42\, 
\mathrm{Tesla}$ and a critical field of $\Delta_c = 0.482\, \mathrm{Tesla}$.  
Although the nuclear dynamics will depend on the 
specific values for the coupling constants, we show below 
that our key results, namely the near-reversal of nuclear dynamics and the 
near-absence 
of electron spin decoherence, are valid regardless of the particular 
values of the hyperfine coupling constants.  Furthermore, we find that 
these results show little dependence on the number of nuclei $N$.

First, we verify that even at high fields nuclear dynamics can occur 
in the form of electron mediated flip-flop between nuclei.
The primary observable we use to 
quantify nuclear spin dynamics is the overlap of the nuclear state at time $t$ 
with the initial nuclear state 
$\ket{\v{z}_0}$, which is specified by the projection operator
\begin{equation}
P_{\v{z}_0} = \ket{\v{z}_0}\bra{\v{z}_0}.
\end{equation}
In the absence of nuclear flip-flop, this operator will have the value 
$1$ for all times $t$.
Figure 1 shows the free evolution of $P_{\v{z}_0}(t)$ at
$B = 0.030, 0.121, .482, 1.93, 7.72 \,\mathrm{Tesla}$ 
(i.e., $B = \frac{1}{16}\Delta_c, \frac{1}{4}\Delta_c, \Delta_c, 
4\Delta_c, 16\Delta_c$) given the initial state $\ket{\psi_0}$ 
defined in \refeqn{IC1}.  We see clearly that above the critical field, 
the magnitude of 
nuclear dynamics decreases as the external field increases, because 
the coupling between nuclear states due to $V_{\mathrm{eff}}$ scales 
as $1/B$.  In essence, increasing the magnetic field acts to 
gradually ``freeze out" nuclear flip-flop. 

Figure 2 shows the magnitude of $\ensavg{S_+(t)}$ for the 
same system.  At low fields, where extensive nuclear dynamics occur, the 
magnitude of $\ensavg{S_+(t)}$ decays 
because the electron experiences a fluctuating magnetic field.  This 
effect is very 
similar to what is observed in nuclear spectral 
diffusion\cite{deSousa:03B1}, except that here
the coupling constant between nuclei depends on the external magnetic 
field.  As we begin to freeze 
out nuclear flip-flops at higher fields, the magnitude of $\ensavg{S_+(t)}$ 
does not decay substantially from its initial value because the electron 
simply precesses at some frequency governed by the initial nuclear 
configuration and its effective Overhauser field.

Although these simulations are useful to explore the origin of nuclear 
dynamics, it is important to note that in real experiments the magnitude 
of the in-plane magnetization of the electron, $\ensavg{S_+(t)}$, will 
decay even in the absence of nuclear flip-flop.  In a real system the 
initial state is unlikely to be a $z$-polarized state of the nuclei, 
or even a pure state of the nuclei; rather, the initial nuclear 
density matrix will likely be highly mixed.  In other words, the 
initial density matrix will contain incoherent contributions from a 
variety of initial nuclear states.  Even if the \emph{magnitude} of 
$\ensavg{S_+}$ undergoes no decay for each of these initial nuclear 
states, the frequency 
of precession for each initial state will be different, leading to 
overall dephasing.  This dephasing is what is 
normally known as ``inhomogeneous broadening"; it is not necessarily 
intrinsic to the system but is only due to the mixed initial state.  
Thus, it becomes necessary to use the spin echo experiment to remove 
this inhomogeneous decay and to recover the intrinsic decay constant 
$T_2$.  
To investigate this effect, we will now use a fully mixed nuclear density 
matrix as an initial state (see \refeqn{Rho0Def}).

First, we confirm that a mixed initial state does indeed 
lead to inhomogeneous broadening in our system.  
Figure 3 shows that, as expected, 
the ensemble of Overhauser fields experienced by the electron due to 
the mixed initial nuclear state leads to dephasing on a timescale 
that is 
nearly independent of the external field.  
We also see that at high fields, the mixed initial state leads to a faster 
decay of $\abs{\ensavg{S_+(t)}}$ than the pure initial state, due to the 
presence of both intrinsic and inhomogeneous broadening ($T_2^* < T_2$).  

We now consider the dynamics of this system under the spin echo experiment. 
As discussed abve, because $H$ does not have the form given in \refeqn{HE}, we do 
not necessarily expect the spin echo experiment to remove all dephasing.  
Figures 1 and 2 confirmed that for a pure initial state at $B = 0.482\, 
\mathrm{Tesla}$, this system undergoes both 
nuclear flip-flop and in-plane magnetization decay.  However, Figure 4 
demonstrates that if we now perform a spin echo experiment, we obtain 
the remarkable result that when $B \geq \Delta_c$, nearly the full 
magnitude of the in-plane magnetization is recovered by the spin echo 
pulse sequence.  It might be expected that when nuclear flip-flop is 
``frozen out" at very high fields (for instance, at $B = 7.72
\,\mathrm{Tesla}$), the spin echo experiment will reverse 
nearly all in-plane magnetization decay.  However, it is surprising
that in systems at much lower fields, for which more complex nuclear 
dynamics occur, the same reversal of decay 
takes place.  In fact, nuclear flip-flop dynamics due to 
dipolar nuclear coupling are known to be responsible for 
irreversible nuclear spectral diffusion 
\cite{deSousa:03B1,deSousa:04X}.  Yet the simulation in 
Figure 4 shows that in contrast to this known behavior for dipolar 
couplings, \emph{the analogous nuclear dynamics induced by the hyperfine 
interaction do not lead to irreversible spin echo envelope decay, except for a 
small visibility loss.}  

To probe this effect, we can directly examine the operators 
$S_+(\tau)$ and $\tilde{S}_+(\tau';\tau)$.  
Figures 5a,b, and c show the matrix 
representation of the operator $S_+(\tau)$ at $\tau 
= 0,.1, 100000 \, \mathrm{ns}$, for the block connecting electron 
spin down to electron spin up states.  At $t=0$, the operator $S_+(0)$ 
acts as the identity on the nuclear states and there are no off-diagonal 
contributions.  However, as time evolves, 
nuclear dynamics, which lead to dephasing, can be clearly seen in the 
non-trivial action of $S_+(\tau)$ on the nuclear states, as evidenced by the 
increasing off-diagonal structure in Figures 5a-c.  Next, a 
$\pi$-pulse is applied at $\tau = 100000\,\mathrm{ns}$.  Figures 
5d and e then show the 
matrix representation of the operator 
$\tilde{S}_+(\tau';100000\,\mathrm{ns})$ at $\tau' = 99999.9,100000 \, 
\mathrm{ns}$.  The evolution of the operator runs backwards after 
the application of the $\pi$-pulse, so that at $\tau' = 100000\, 
\mathrm{ns}$, we 
find that $\tilde{S}_+(\tau';100000\,\mathrm{ns})$ is nearly equal to 
$S_+(0)$.  Because the operator 
$\tilde{S}_+(\tau;\tau)$ is nearly the identity 
operator with repect to the nuclear states, the in-plane magnetization 
will remain close to its initial value, regardless of the initial state.

We now examine the importance of the specific values of the hyperfine 
constants, $A_j$, for this nuclear dynamics reversal to occur.  So far, 
we have offered no justification for our particular selection of 
hyperfine constants, which 
were chosen at random from a uniform distribution.  Because the 
nuclear and electronic dynamics depend on the exact value of the hyperfine 
constants, the absence of spin decay that we observe might be highly 
system dependent.  To probe this issue, we will evaluate the magnitude of 
the spin-echo decay as a function of $B$ for a variety of system parameters 
and sizes.  For each system, we select hyperfine constants from the 
uniform distribution on the interval $\left[0,.6\right]\, 
\mathrm{Tesla}$ and evaluate the time-averaged magnitude of the 
spin-echo envelope or the \emph{visibility loss}, $v$ , where
\begin{equation} \label{Equation::Visibility}
v = 1/2 - \overline{{\abs{\ensavg{\tilde{S}_+(\tau;\tau)}}}}/\hbar.
\end{equation}
In \refeqn{Visibility} the time average is taken over some suitably 
long time interval (approximately $500 \mathrm{ns}$ for our systems).  
Figure 6 shows a graph of the results for 
several values of $N$, where we have scaled our results to the
critical field of each system (because the $A_j$'s were selected 
randomly, the critical field for each system varies).  These plots 
clearly demonstrate a universal scaling of visibility (i.e. the 
magnitude of the spin echo envelope) with magnetic field, regardless of 
the particular values of the hyperfine coupling constants.  Instead, the 
visibility of a given system depends only its critical field 
$\Delta_c$.  Furthermore, we see that this behavior is independent of 
system size; at external fields greater than $\Delta_c$, the visibility 
loss scales as $(\Delta_c/B)^2$ for every value of $N$ simulated.  
That this time-reversal phenomenon seems to 
be completely independent of the specific values of the hyperfine 
constants $A_j$ and of the number of nuclei is remarkable, 
considering that the nuclear dynamics themselves are very sensitive to the 
particular values of $A_j$.

\section{Discussion and Conclusions}
\label{Section::Conclusions}
We have studied here the dynamics of a system of one 
electron interacting with $N$ nuclei via the hyperfine interaction with exact 
diagonalization methods.  We have found that even at external magnetic fields 
above the critical field $\Delta_c$, nuclear dynamics can still 
occur as a result of the second-order coupling between nuclei induced by the 
electron-nuclear hyperfine interaction.  These dynamics cause
the electron to experience a fluctuating Overhauser field, giving rise to decoherence
similar to the effect of dipolar nuclear spectral diffusion.  
However, unlike nuclear 
spectral diffusion, these flip-flop dynamics due to the hyperfine 
interaction can be nearly completely reversed using a single spin echo pulse 
sequence.  This reversal of nuclear dynamics in turn reverses the decay of 
the in-plane electron spin magnetization, leading to a negligible decay of 
the spin echo envelope function.  The latter appears to be better characterized 
as a visibility loss\cite{Yablonovitch:03}.  Finally, we have found that this loss 
of visibility obeys a universal scaling with the external magnetic field that 
depends only on the critical field of the system.

It should be noted that the entanglement generated by the hyperfine 
interaction between the electron and nuclei 
has already been studied at zero external field,
where it was concluded that it will be necessary to 
remove or at least to minimize this entanglement in a quantum computer 
implementation based on electron spins in solids
\cite{Schliemann:02B}.  Our investigation shows that the 
spin echo pulse sequence accomplishes precisely this task by reversing the 
dynamics of the system.  Furthermore, we demonstrate that this 
reversal is possible at all field strengths above $\Delta_c$, apart from a 
small loss of visibility on the order of $(\Delta_c / B)^2$.

Several points merit further discussion.  The first is the impact of the 
observed universal scaling on the so-called ``visibility problem" 
discussed by Yablonovitch et al. \cite{Yablonovitch:03}.  In that paper, 
the authors point out that even small scale fluctuations (i.e. loss of 
visibility) in the spin-echo signal caused by the electron-nucleus hyperfine 
interaction can be fatal to quantum computation if they are above the error 
threshhold ($10^{-4} - 10^{-6}$).  Thus, even if the spin echo 
experiment recovers \emph{most} of the in-plane coherence, it is 
important to know how large the external magnetic field must be before 
the loss of in-plane coherence is below the error threshhold for quantum 
computation.  Because the critical field is shown here to scale approximately 
as $\Delta_c = B_c / \sqrt{N}$, for large numbers of nuclei the critical field will be 
substantially lower than the total Overhauser field of the nuclei.  For 
instance, the total Overhauser field of an electronic impurity in 
GaAs is approximately $B_c = 2.24\, \mathrm{Tesla}$; however, the 
critical field is only $\Delta_c = 1.25\times10^{-3}\, 
\mathrm{Tesla},$ since there are approximately $N=10^6$ 
nuclei interacting with the electron\cite{Shenvi:04X1}.  Hence, an 
external field of $B = 1\,\mathrm{Tesla}$ would lead to a spin echo 
visibility loss on the order of $10^{-6}$ and for any field larger than this
would therefore be sufficient to allow fault tolerant quantum computation (in the absence of other 
decoherence mechanisms).  

Second, we should point out the relationship between spin coherence recovery 
and entanglement.  \refeqto{Heffp}{Veffm} show that spin-up states and 
spin-down states evolve via different effective Hamiltonians.  Because the 
evolution of the nuclear spins depends on the state of the electron spin, the 
nuclei become entangled with the electron, resulting in a loss of electron 
spin coherence.  As we have shown, the spin echo sequence reverses this loss 
of entanglement almost completely.  In other words, the effect of the spin 
echo sequence can be understood as disentangling the electron spin from the 
nuclear spins.  That the hyperfine interaction should generate entanglement 
is well-known \cite{Schliemann:02B}; what is interesting and new from the current
study is that the spin echo 
sequence should be able to so effectively remove this entanglement.  

Finally, we turn to the question of the origin of the observed universal 
scaling.  It is our belief that such a universal scaling which is 
invariant to both system size and the choice of coupling constants must 
result from a hidden near symmetry in the system.  A hidden symmetry might 
also account for the oscillatory behavior and long timescale
persistence of spin correlation functions observed in \cite{Erlingsson:04}.
This claim can perhaps be better understood by an analogy.  The absence of 
longitudinal decay at high magnetic fields can be thought of as a result 
of the near-commutation of $H$ and $S_z$.  As a result of this 
near-commutation, $S_z$ is a ``almost" a good quantum number for the 
system and the longitudinal component of the electron spin decays only a 
small amount.  In this paper, we have observed that the in-plane component 
of the electron spin decays only a small amount under the spin echo 
experiment.  This fact suggests that there might be some operator 
which nearly commutes with $H$, leading to a similar suppression of decay.  
The difference between the two cases is that $T_1$ suppression is linked 
to the energy gap between up and down electron spin states.  There is no 
obvious analogous energy gap for $T_2$ processes, because these processes 
conserve electron spin.  It is known that the hyperfine Hamiltonian in 
\refeqn{HFull} commutes with a set of operators discovered by Gaudin 
\cite{Gaudin:76P, Schliemann:03P}.  However, this fact does not 
immediately explain the presence of the observed universal scaling, nor 
the near-complete spin echo reversal.

Much future work remains to be done on this problem.  The most obvious 
question that remains unanswered is the identity of the
conjectured hidden symmetry and the mechanism by which it suppresses spin 
echo decay.  This symmetry might be the one mentioned above, or it might 
be one that is completely new.  In either case, determining its effect 
on the spin echo envelope would provide fascinating insight into the 
nature of spin dynamics.  

In conclusion, we provide numerical evidence that a spin echo sequence 
is able to remove a substantial part of the hyperfine induced entanglement 
of a single electron spin interacting with a bath of nuclear spins.  At 
high magnetic fields ($B \gg \Delta_c$) this residual entanglement reveals 
itself as a universal visibility loss, which is shown to depend only on 
the $(\Delta_c / B)$ ratio (see \refeqn{BCPrimeDef}).  

Ackowledgements.  We acknowledge support from the DARPA SPINS program and 
ONR under grant No. FDN001-01-1-0826.

\pagebreak[4] 
\begin{figure}
\includegraphics{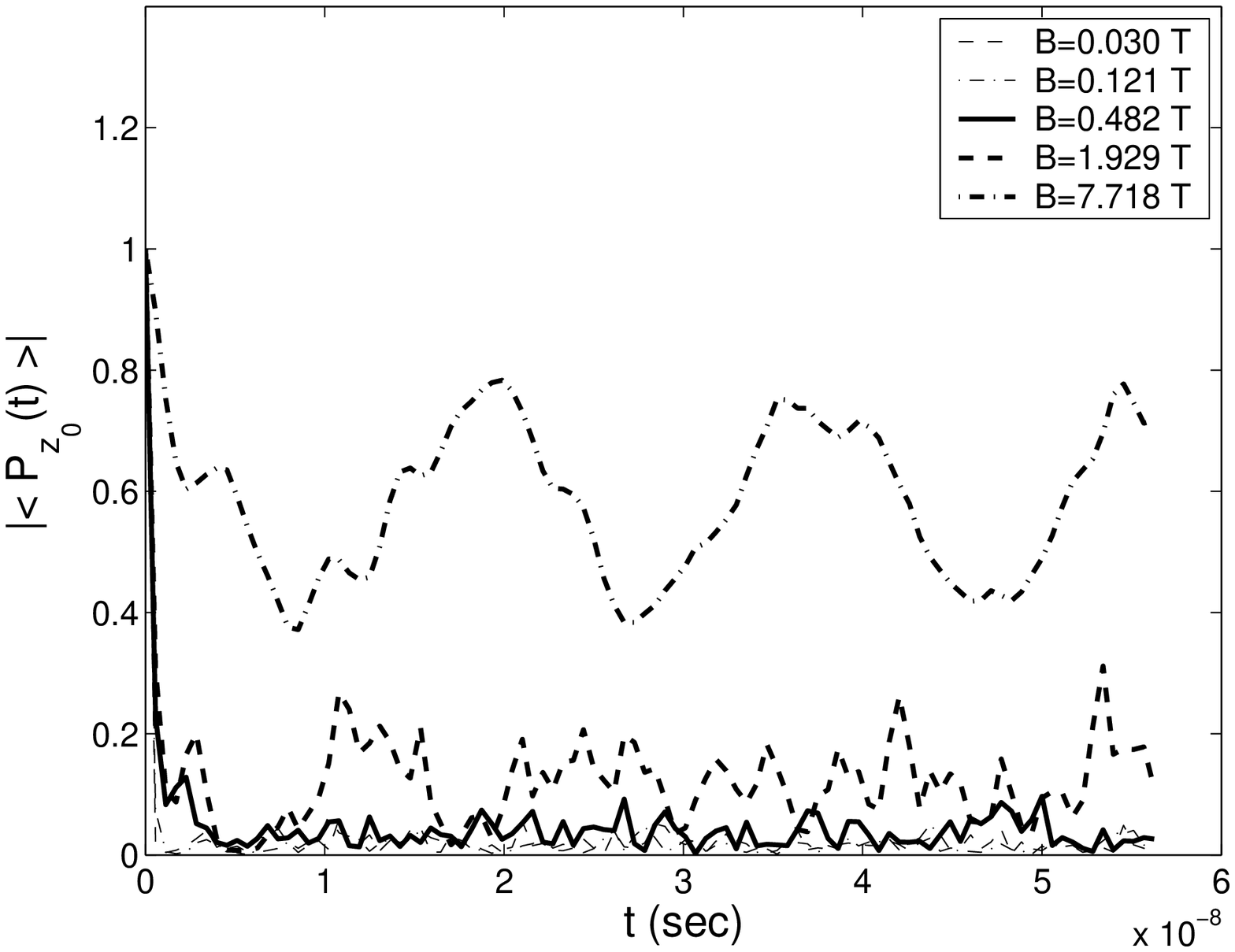}
\caption{
The expectation value of $P_{\v{z}_0} = 
\id_S\tensorm\ket{\v{z}_0}\bra{\v{z}_0}$ as a 
function of $t$ at $B = 0.030, 0.121, .482, 1.93, 7.72 
\,\mathrm{Tesla}$ for the initial state $\ket{\psi_0}$ (see 
\refeqn{IC1}).  In this 
example $N=9$ and $\Delta_c = .482\, \mathrm{Tesla}$.  As the external field 
increases, nuclear dynamics is slowly ``frozen out".  However, nuclear 
dynamics clearly persist well above the critical field.  
}
\end{figure}

\pagebreak[4] 
\begin{figure}
\includegraphics{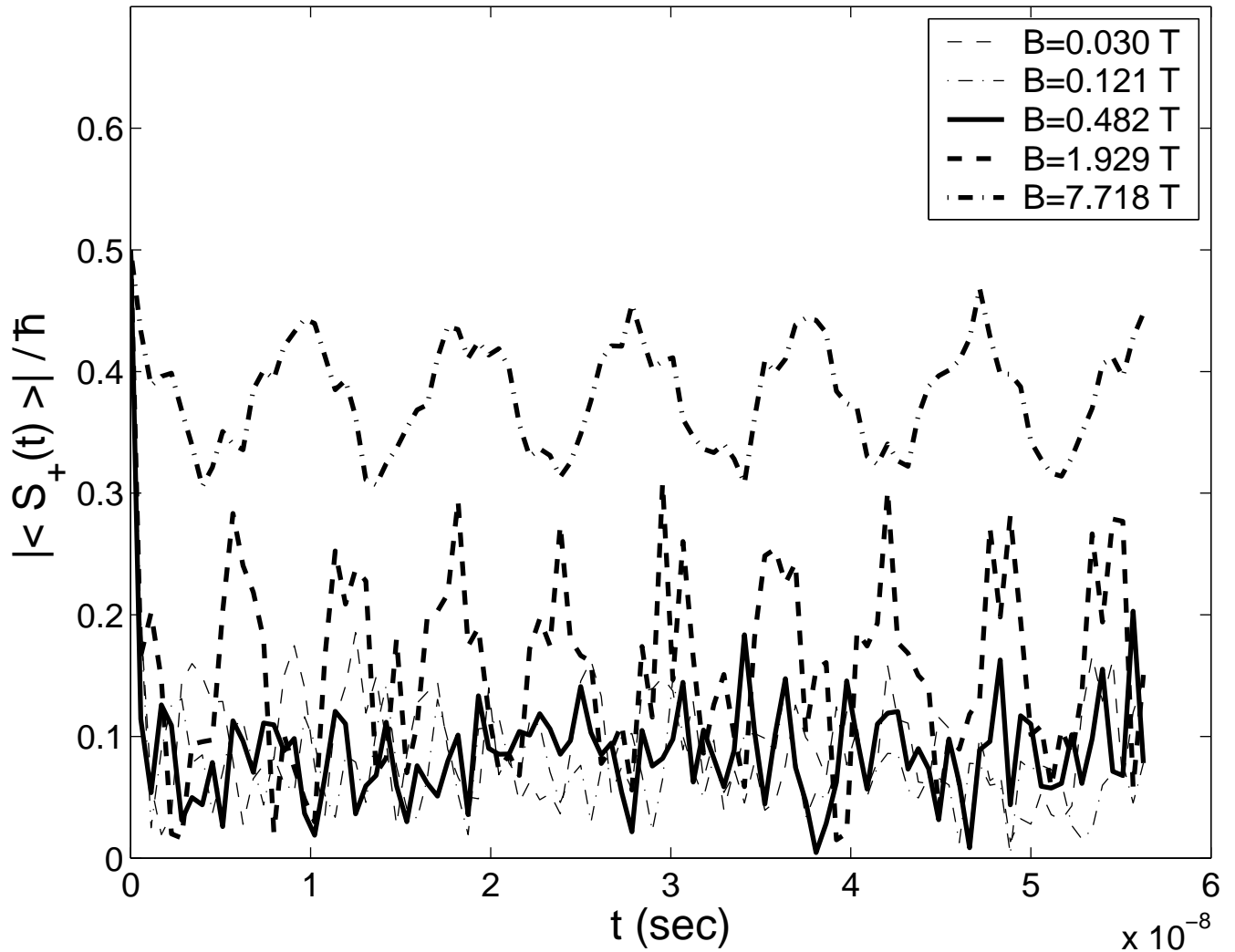}
\caption{
The expectation value of the magnitude of the in-plane magnetization 
$\abs{\ensavg{S_+(t)}}$ as a 
function of $t$ at $B = 0.030, 0.121, .482, 1.93, 7.72 
\,\mathrm{Tesla}$ for the same system as in Figure 1, i.e. the initial 
state is again the pure state $\ket{\psi_0}$.  As 
the nuclear dynamics is slowly ``frozen out", there is a corresponding 
decrease in the magnitude of in-plane magnetization decay.  
}
\end{figure}

\pagebreak[4] 
\begin{figure}
\includegraphics{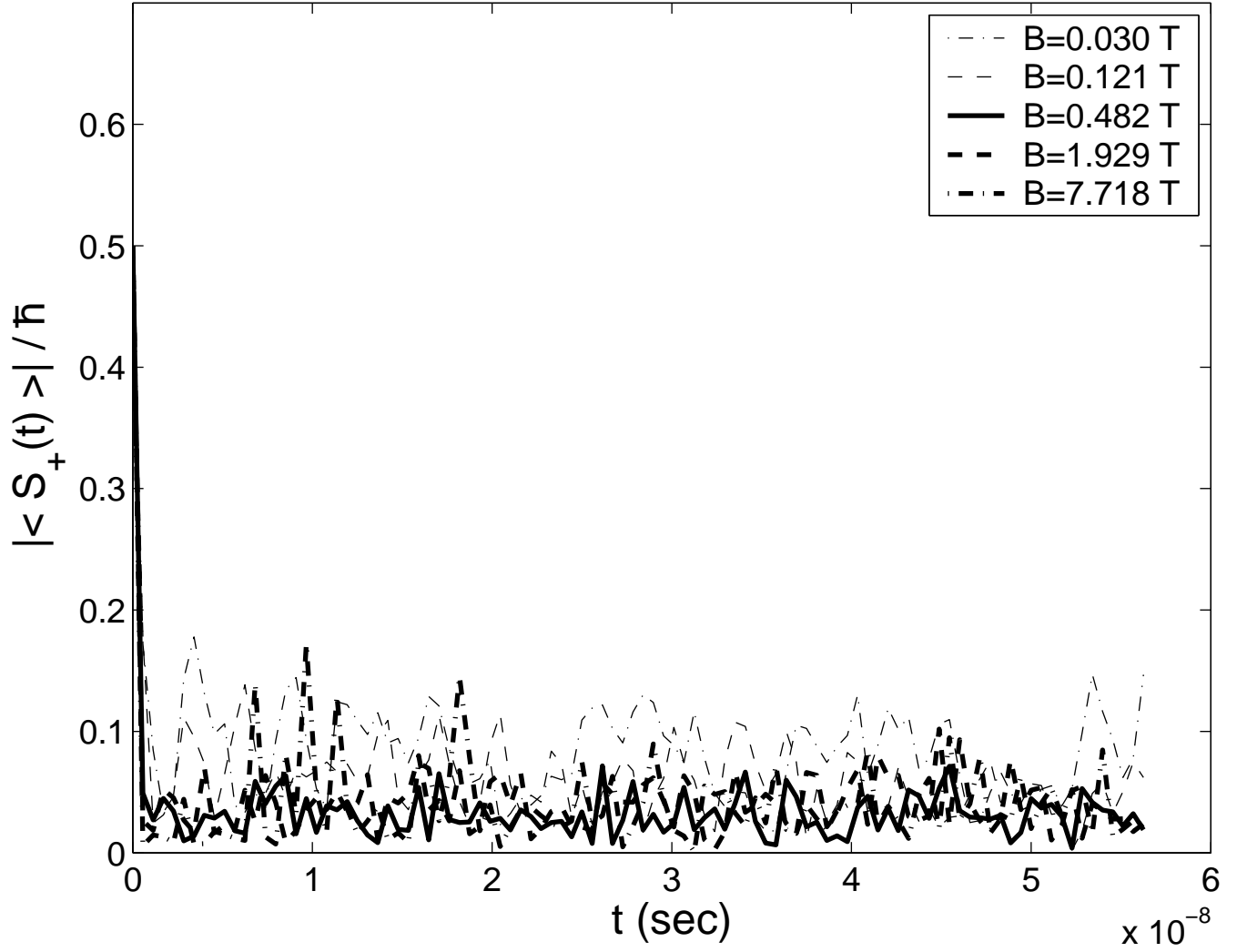}
\caption{
The expectation value of the magnitude of the in-plane magnetization 
$\abs{\ensavg{S_+(t)}}$ as a function of $t$ at $B = 0.030, 0.121, .482, 
1.93, 7.72 \,\mathrm{Tesla}$ for an
initial density matrix $\rho(0)$ corresponding to a completely mixed 
nuclear state (see \refeqn{Rho0Def}).  Inhomogeneous broadening causes the 
in-plane magnetization 
to decay on approximately the same timescale ($T_2^*$) independent of the 
external field.}
\end{figure}

\pagebreak[4] 
\begin{figure}
\includegraphics{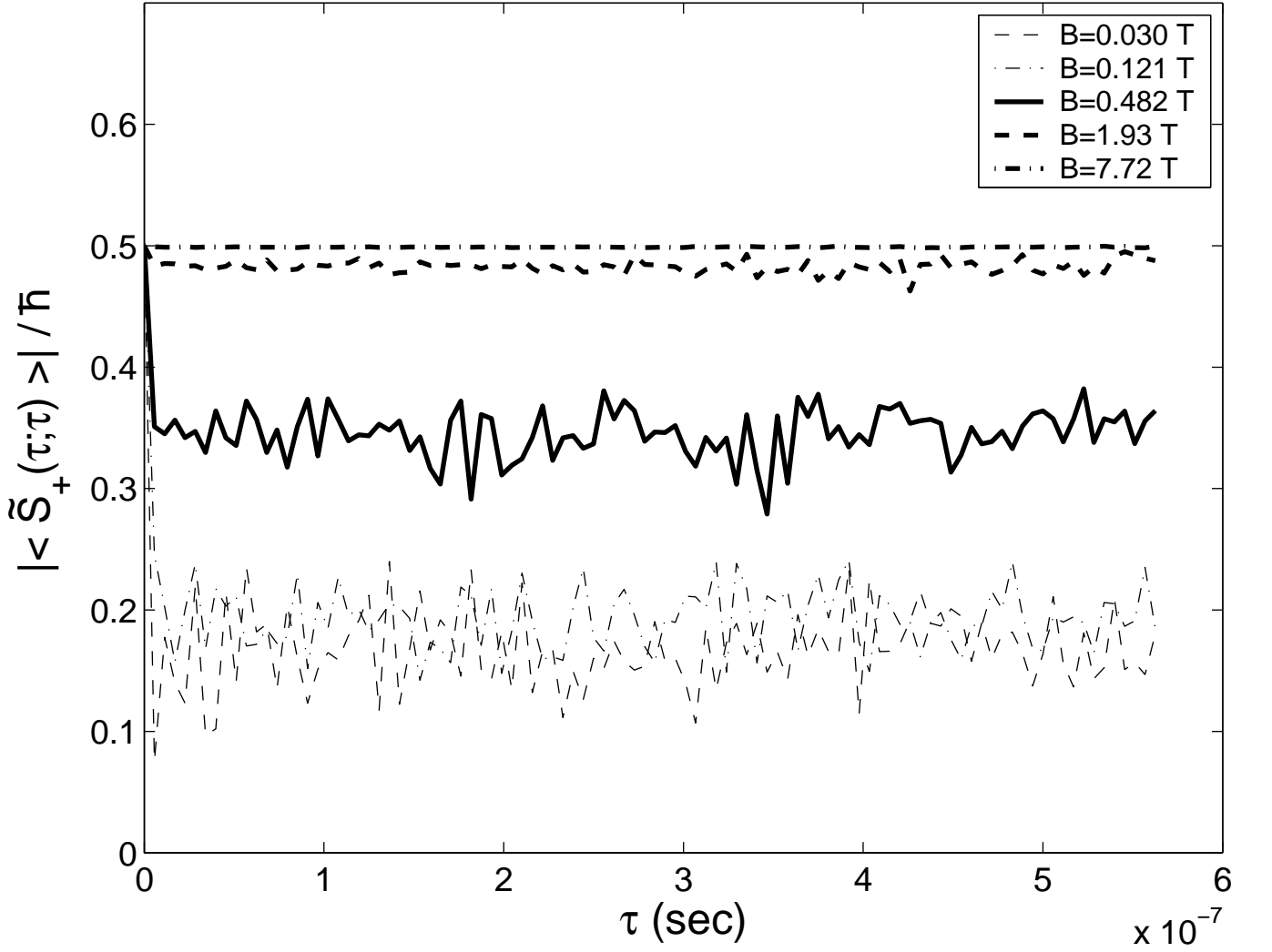}
\caption{
The expectation value of the magnitude of the spin echo 
envelope $\abs{\ensavg{\tilde{S}_+(\tau;\tau)}}$ as a 
function of $\tau$ at $B = 0.030, 0.121, .482, 
1.93, 7.72 \,\mathrm{Tesla}$.  The initial nuclear state is again 
the mixed state specified by $\rho(0)$ (see \refeqn{Rho0Def}).
Above the critical field $\Delta_c$, the spin echo experiment removes 
nearly all decay of the in-plane magnetization, even for systems 
displaying substantial nuclear dynamics (i.e. $B = 1.93\, 
\mathrm{Tesla}$).
Note that this effect persists even at long times.
}
\end{figure}

\pagebreak[4] 
\begin{figure}
\includegraphics[width=7in]{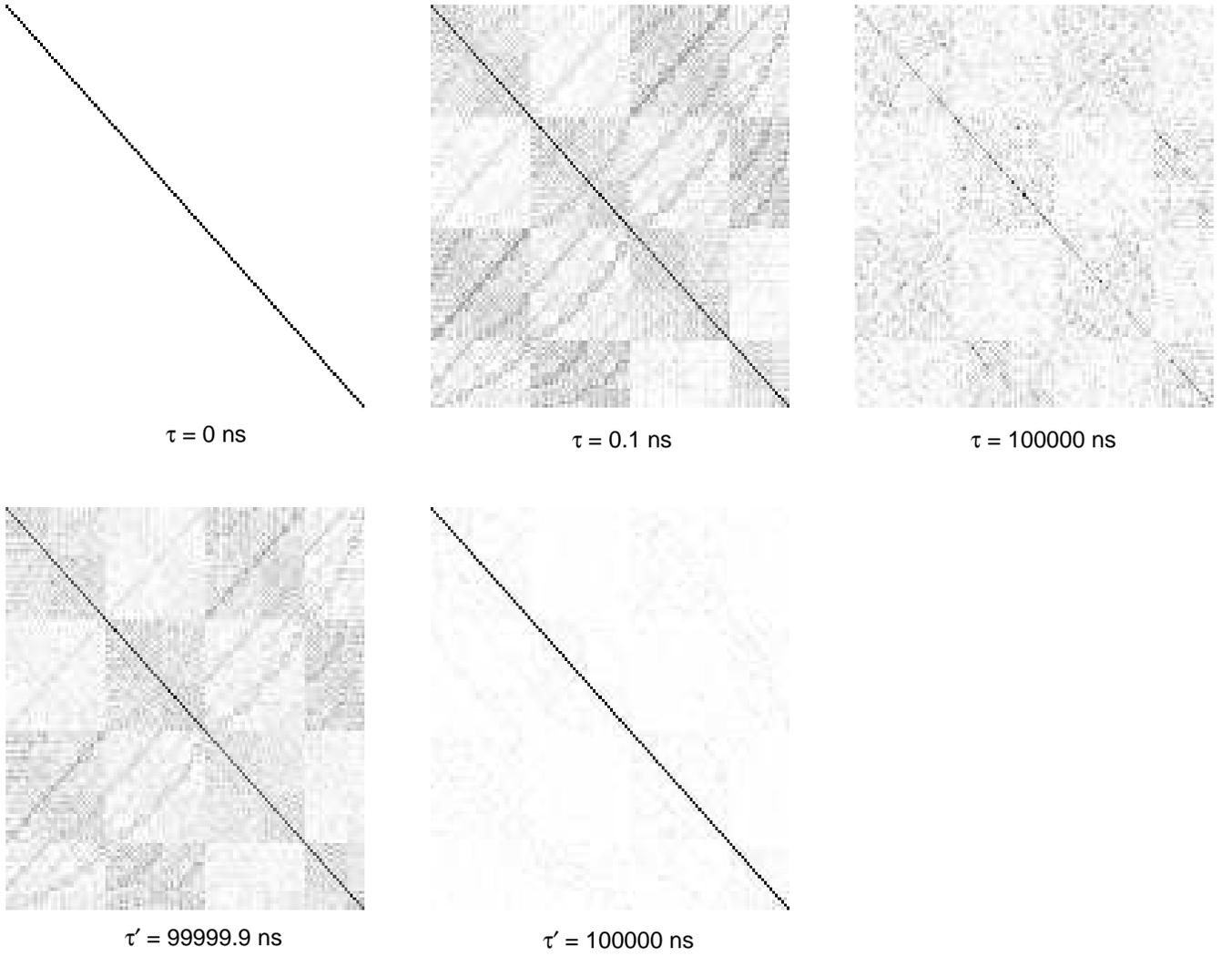}
\caption{
The matrix representation of the operator 
$\tilde{S}_+(\tau';\tau)$ at $B = \Delta_c = 0.482\, \mathrm{Tesla}$ over the course of a spin echo experiment where
a $\pi$-pulse is applied at $\tau = 100000\, \mathrm{ns}$.  The operators are shown in the 
$z$-basis representation of the nuclear states; the intensity of each 
point on the plot represents 
the amplitude of the corresponding matrix element between nuclear states.  
For clarity, we consider only the primary 
contributing block of the operator (i.e. the block connecting electron 
spin down to electron spin up states) as the other blocks are much 
smaller in magnitude.  
At short times, nuclear dynamics are negligible.  At longer times, nuclear 
dynamics are substantial, leading to potential decay of 
$S_+$.  However, the final panel shows that the spin echo experiment 
nearly reverses all nuclear dynamics, leading to a recovery of in-plane 
magnetization.
}
\end{figure}

\pagebreak[4] 
\begin{figure}
\includegraphics[width=7in]{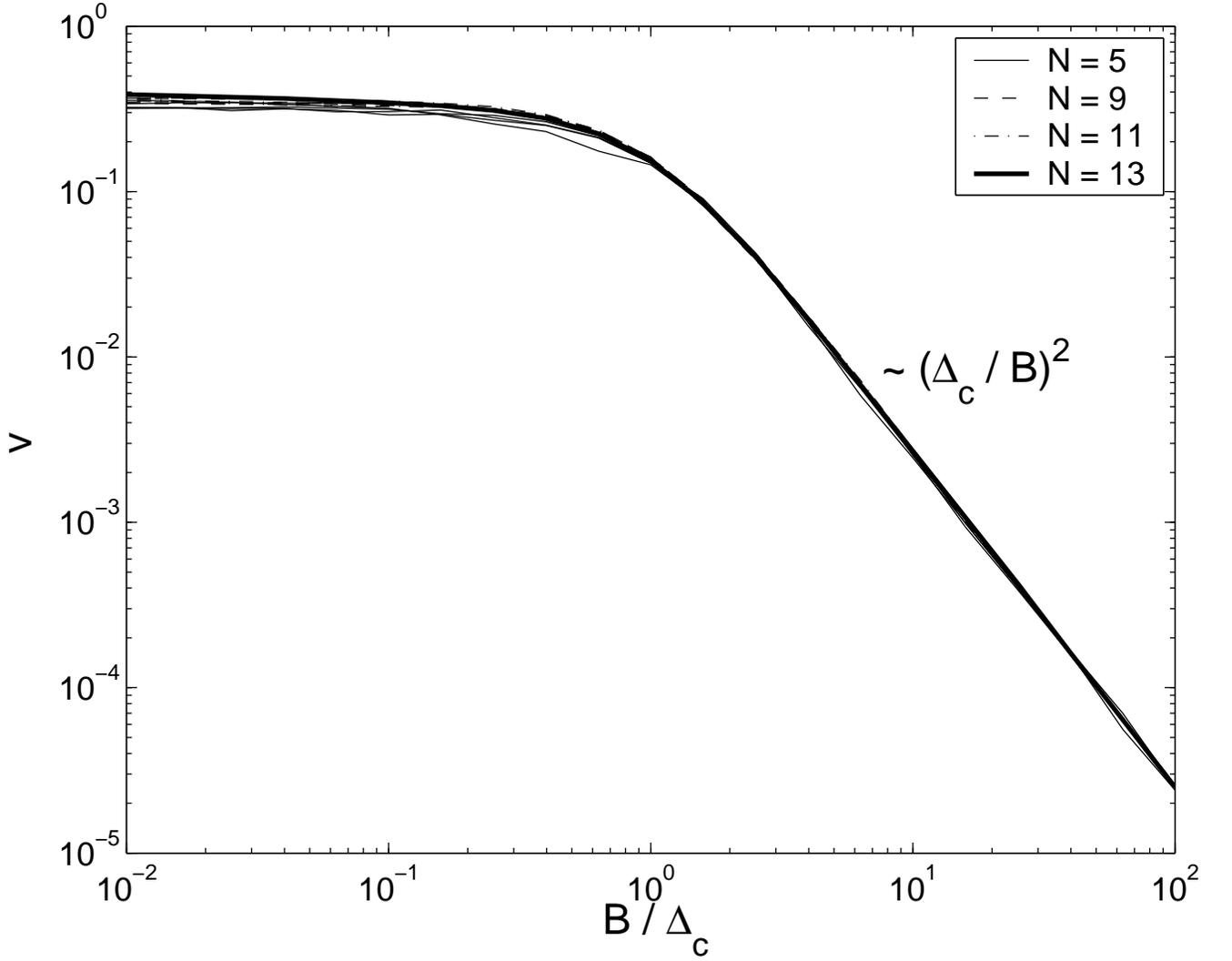}
\caption{
The universal scaling of visibility loss, $v$, (see 
\refeqn{Visibility}) 
versus external field, $B / \Delta_c$, evaluated for systems of $N=5, N=9, 
N=11, N=13$ with randomly generated sets of hyperfine coupling 
contstants.  Five systems were simulated for the system sizes $N=5, 
N=9,$ and $N=11$.  Only one system was simulated 
for $N=13$ because the large size of the system required substantial 
computation time (approximately five days on a multiple-node 
workstation).  Above the critical field $B > \Delta_c$, the visibility 
loss scales as $(\Delta_c / B)^2$, independent of the selection of 
the hyperfine constant values and the size of the system, $N$.}
\end{figure}



\end{document}